\documentstyle[sprocl]{article}

\def\NPB{{\em Nucl. Phys.} B}
\def\PLB{{\em Phys. Lett.}  B}
\def\PRL{\em Phys. Rev. Lett.}
\def\PRD{{\em Phys. Rev.} D}

\def\ra{\rightarrow}

\def\J{{J/\psi|}}
\def\ov{\overline}
\def\be{\begin{eqnarray}}
\def\en{\end{eqnarray}}
\def\non{\nonumber}
\def\la{\langle}
\def\ra{\rangle}

\def\J{{J/\psi}}
\def\ov{\overline}
\def\Lqcd{{\Lambda_{\rm QCD}}}

\begin{document}

\title{PHENOMENOLOGICAL APPLICATIONS OF QCD
FACTORIZATION TO SEMI-INCLUSIVE $B$ DECAYS}

\author{HAI-YANG CHENG}

\address{Institute of Physics, Academia Sinica, Taipei, Taiwan
115, R.O.C.\\ E-mail: phcheng@ccvax.sinica.edu.tw}


\maketitle\abstracts{We have systematically investigated the
semi-inclusive $B$ decays $B\to MX$, which are manifestations of
the quark decay $b \to Mq$, within a framework inspired by
QCD-improved factorization. These decays are theoretically clean
and have distinctive experimental signatures. We focus on a class
of these that do not require any form factor information and
therefore may be especially suitable for extracting information on
the angles $\alpha$ and $\gamma$ of the unitarity triangle.  The
strong phase coming from final-state rescattering due to hard
gluon exchange between the final states can induce large rate
asymmetries for tree-dominated color-suppressed modes
$(\pi^0,\rho^0,\omega)X_{\bar s}$. The nonfactorizable hard
spectator interactions in the 3-body decay $B\to Mq_1\bar q_2$,
though phase-space suppressed, are extremely important for the
tree-dominated modes $(\pi^0,\rho^0,\omega)X_{\bar s},~\phi X$,
$\J X_s,\J X$ and the penguin-dominated mode $\omega X_{s\bar s}$.
Our result for ${\cal B}(B\to \J X_s)$ is in agreement with
experiment. $\ov B^0_s\to (\pi^0,\rho^0,\omega)X_{\bar s}$,
$\rho^0X_{s\bar s}$, $\ov B^0\to(K^-X,K^{*-}X)$ and $B^-\to
(K^0X_s,K^{*0}X_s)$ are the most promising ones in searching for
direct CP violation: they have branching ratios of order
$10^{-6}-10^{-4}$ and CP rate asymmetries of order $(10-40)\%$. }

\section{Why Semi-inclusive $B$ Decays ?}
The semi-inclusive decays $B\to M+X$ that are of special interest
originate from the quark level decay, $b \rightarrow M + q$. They
are theoretically cleaner compared to exclusive decays and have
distinctive experimental signatures \cite{Soni,Browder}. The
theoretical advantages are : (i) A very important theoretical
simplification occurs in the semi-inclusive decays over the
exclusive decays if we focus on final states such that $M$ does
not contain the spectator quark of the decaying $B(B_s)$ meson as
then we completely by-pass the need for the transition form factor
for $B(B_s) \to M$. Recall that for the exclusive case, in
general, we need a knowledge of two such form factors if $M$ is a
pseudoscalar meson or of four form factors if $M$ is a vector
meson. (ii) There is no troublesome infrared divergent problem
occurred at endpoints when working in QCD factorization, contrary
to the exclusive decays where endpoint infrared divergences
usually occur at twist-3 level. (iii) Some unknown strong (hard
and soft) phases may arise from final-state interactions. However,
these phases are mostly washed out in semi-inclusive decays.
Consequently, the predictions of the branching ratios and partial
rate asymmetries for $B\to MX$ are considerably clean and
reliable. Since these semi-inclusive decays also tend to have
appreciably larger branching ratios compared to their exclusive
counterparts, they may therefore be better suited for extracting
CKM-angles and for testing the Standard Model.

Earlier studies of semi-inclusive decays are based on generalized
factorization \cite{Browder} in which nonfactorizable effects are
treated in a phenomenological way by assuming that the number of
colors $N_c^{\rm eff}$ is a free parameter to be fitted to the
data or naive factorization \cite{Soni} where $N_c^{\rm eff}=3$.
Apart from the unknown nonfactorizable corrections, the
factorization approach encounters another major theoretical
uncertainty, namely the gluon's virtuality $k^2$ in the penguin
diagram is basically unknown, rendering the predictions of CP
asymmetries not trustworthy.

The aforementioned difficulties with the conventional methods can
be circumvented in the BBNS (Beneke, Buchalla, Neubert, Sachrajda)
approach of QCD-improved factorization \cite{BBNS1}. Recently QCD
factorization has been applied to charmless semi-inclusive decays
$B\to K(K^*)X$ and $B\to\phi X_s$ in \cite{He,Hephi}. Our goal is
to extend the application of BBNS idea of QCD factorization to a
certain class of semi-inclusive decays. In this regard our
approach complements the recent works of He {\it et al}
\cite{He,Hephi}.

\section{QCD Factorization}
We wish to suggest that the idea of QCD factorization \cite{BBNS1}
can be extended to the case of semi-inclusive decays, $B \to M +
X$, with rather energetic meson $M$, say $E_M \geq 2.1~{\rm GeV}$.
Recall that it has been shown explicitly \cite{BBNS1} that if the
emitted meson $M_2$ is a light meson or a quarkonium in the
two-body exclusive decay $B\to M_1M_2$ with $M_1$ being a recoiled
meson, the transition matrix element of an operator $O$, namely
$\la M_1M_2|O|B\ra$, is factorizable in the heavy quark limit.
Schematically one has \cite{BBNS1}
 \be
 \la M_1M_2|O_i|B\ra &=& \la M_1M_2|O_i|B\ra_{\rm fact}
 \left[1+\sum r_n\alpha_s^n+{\cal O}({\Lqcd\over
m_b})\right]  \non \\ &=& \sum_jF_j^{BM_1}(m_2^2)\int^1_0 du\,
T_{ij}^I(u)\Phi_{M_2}(u) \non
\\ && +\int^1_0 d\xi \,du\,dv
\,T^{II}_i(\xi,u,v)\Phi_B(\xi)\Phi_{M_1}(u)\Phi_{M_2}(v),
\label{qcdf}
 \en
where $F^{BM_1}$ is a $B-M_1$ transition form factor, $\Phi_M$ is
the light-cone distribution amplitude, and $T^I,~T^{II}$ are
perturbatively calculable hard scattering kernels.  The second
hard scattering function $T^{II}$, which describes hard spectator
interactions, survives in the heavy quark limit when both $M_1$
and $M_2$ are light or when $M_1$ is light and $M_2$ is a
quarkonium \cite{BBNS1}. The factorization formula (\ref{qcdf})
implies that naive factorization is recovered in the
$m_b\to\infty$ limit and in the absence of QCD corrections.
Nonfactorizable corrections are calculable since only hard
interactions between the $(BM_1)$ system and $M_2$ survive in the
heavy quark limit.

In order to have a reliable study of semi-inclusive decays both
theoretically and experimentally, we will impose two cuts
\cite{CS}. First, a momentum cutoff imposed on the emitted light
meson $M$, say $p_M>2.1$ GeV, is necessary in order to reduce
contamination from the unwanted background and ensure the
relevance of the two-body quark decay $b\to Mq$. For example, an
excess of $K(K^*)$ production in the high momentum region, $2.1<
p_{K(K^*)}<2.7$ GeV, will ensure that the decay $B\to K(K^*)X$ is
not contaminated by the background $b\to c$ transitions manifested
as $B\to D(D^*)X\to K(K^*)X'$ and that it is dominated by the
quasi two-body decay $b\to K(K^*)q$ induced from the penguin
process $b\to sg^*\to sq\bar q$ and the tree process $b\to u\bar u
s$. Second, it is required that the meson $M$ does not contain the
spectator quark in the initial $B$ meson and hence there us no
$B-M$ transition form factors. Under these two cuts, we argue that
the factorization formula (\ref{qcdf}) can be generalized to the
semi-inclusive decay:
 \be
 \la MX|O|B\ra &=& \la MX|O|B\ra_{\rm
fact}\left[1+\sum r_n\alpha_s^n+{\cal O}({\Lqcd\over m_b})\right]
\non \\ &=& \int^1_0 du\, T^I(u)\Phi_{M}(u)  +\int^1_0 d\xi \,du
\,T^{II}(\xi,u)\Phi_B(\xi)\Phi_{M}(u). \label{qcdfsemi}
 \en
However, this factorization formula is not as rigorous as the one
(\ref{qcdf}) for the exclusive case, as we shall elucidate on
below.

The factorizable hadronic matrix element $\la MX|O|B\ra$ in
general consists of several terms:
 \be \label{factm.e.}
 \la MX|O|B\ra_{\rm fact} &=& \la M|j_1|0\ra\la X|j_2|B\ra
+\underbrace{\la X|j_1'|0\ra\la M|j_2'|B\ra} \non \\
&& {\rm \hskip3.0cm prohibited ~by~2nd~cut} \non \\ &+&
\underbrace{\la X_1M|j_1|0\ra\la X_1'|j_2|B\ra+\la
X_2|j_1'|0\ra\la X_2'M|j_2'|B\ra} \non \\
&& {\rm  not~favored~by~1st~cut,~also}~\alpha_s~{\rm suppressed} \non \\
&+& \underbrace{\la MX|j_1|0\ra\la 0|j_2|B\ra}. \non \\
&& {\rm power~suppressed}
 \en
However, several terms are prohibited or suppressed by
aforementioned two cuts.  The last term in Eq. (\ref{factm.e.}) is
the annihilation contribution and it is suppressed by order
$\Lqcd/m_b$. In comparing Eq. (\ref{qcdfsemi}) to the exclusive
case Eq. (\ref{qcdf}), a crucial simplification that has occurred
is that the semi-inclusive case does not involve any transition
form factor(s).  Since lack of knowledge of these form factors is
often a serious limitation in quantitative applications, this adds
to the appeal of the semi-inclusive case. Note also that when the
emitted meson $M$ is a light meson or a quarkonium, the
nonfactorizable corrections to naive factorization are infrared
safe in the heavy quark limit and hence calculable.

In contrast to the exclusive case, the parton model implies that
the semi-inclusive decay rate of the $B$ meson can be approximated
by that of the free $b$ quark in the heavy quark limit, namely
$\Gamma(B\to MX)\approx \Gamma(b\to Mq)$. Hence, the hard
spectator interactions in semi-inclusive decays should be
suppressed in the heavy quark limit. As we shall see later, they
are suppressed by powers of $(\Lqcd/m_b)$ at the decay rate level.
However, these interactions will gain large enhancement for
tree-dominated color-suppressed modes. Therefore, we will keep
this term in Eq. (\ref{qcdfsemi}).

To the order ${\cal O}(\alpha_s)$, there are two additional
contributions besides vertex corrections: the bremsstrahlung
process $b\to Mq\,g$ ($g$ being a real gluon) and the process
$b\to Mq\,g^*\to Mqq'\bar q'$. The bremsstrahlung subprocess could
potentially suffer from the infrared divergence. However, the
vertex diagram in which a virtual gluon is attached to $b$ and $q$
quarks is also infrared divergent. This together with the
above-mentioned bremsstrahlung process will lead to a finite and
well-defined correction. Note that for exclusive hadronic decays,
the infrared divergence occurring in the $BM_2$ system ($M_2$
being a recoiled meson) is absorbed into in the $B-M_2$ transition
form factors \cite{BBNS1}. This finite correction is expected to
be small as it is suppressed by a factor of $\alpha_s/\pi\approx
7\%$. Since $b\to Mq\,g$ does not interfere with $b\to Mq$, it can
be counted as an order ${\cal O}(\alpha_s)$ correction. In the
presence of bremsstrahlung and the fragmentation of the
quark-antiquark pair from the gluon, the factorizable
configurations $\la X_1M|j_1|0\ra\la X_1'|j_2|B\ra$ and $\la
X_2|j_1|0\ra\la X_2'M|j_2|B\ra$ with $X_1+X_1'=X$ and
$X_2+X_2'=X$, that will break the factorization structure shown in
the first term in (\ref{qcdfsemi}), are allowed in Eq.
(\ref{factm.e.})\cite{He}. In general, one may argue that these
configurations are suppressed since the momentum cut $p_M>2.1$ GeV
favors the two-body quark decay $b\to Mq$ and low multiplicity for
$X$. However, it is not clear to us how rigorous this argument is.
Therefore, in the present paper we will confine ourselves to
vertex-type and penguin-type corrections as well as hard spectator
interactions so that the factorization formula (\ref{qcdfsemi}) is
applicable to semi-inclusive decays at least as an approximation.

Note that QCD factorization is not applicable to the decay $\ov
B^0\to \pi^0 D^0$  because the emitted meson $D^0$ is heavy so
that it is neither small (with size of order $1/\Lqcd$) nor fast
and cannot be decoupled from the $(B\pi)$ system. Hence, by the
same token as the $\ov B^0\to \pi^0 D^0$ decay, the above QCD
factorization formula is also not applicable to $\ov B^0\to
D^0(\ov D^0)X$.

\section{Two-body Decays of the $\lowercase{b}$ Quark}
A major advantage of studying the quasi-two-body decay of the $b$
quark is that it does not involve the unknown form factors and
hence the theoretical uncertainty is considerably reduced. Hence,
we first study the $CP$-averaged branching ratios and direct
$CP$-violating partial rate asymmetries for some two-body hadronic
$b$ decays of interest. The results are shown in Table I. Compared
to the predictions of branching ratios based on naive
factorization \cite{Soni}, there are three major modifications:
(i) Decay modes $\pi^- u$, $\bar K^0d$, $\bar K^{*0}d$ and $K^-u$
are significantly enhanced owing to the large penguin coefficients
$a_6$ and $a_4$. (ii) The modes $\pi^0d,~\rho^0d,~\omega\,d,~\J
s,~\J d$ with neutral emitted mesons are suppressed relative to
the naive factorization ones due to the smallness of $a_2$. (iii)
The $\phi d$ mode has a smaller rate due to the large cancellation
between $a_3$ and $a_5$. That is, while $\phi d$ is QCD-penguin
dominated in naive factorization, it becomes electroweak-penguin
dominated in QCD factorization.

\begin{table}[ht]
\caption{$CP$-averaged branching ratios and partial-rate
asymmetries for some two-body hadronic $b$ decays. For comparison,
the predicted branching ratios and rate asymmetries (in absolute
values for $\gamma=60^\circ$) based on naive factorization [1] are
given in parentheses.}
\begin{center}
\begin{tabular}{ l l c  }
 Mode & BR & PRA(\%)  \\ \hline
 $b\to \pi^-u$ & $1.5\times 10^{-4}~(1.3\times 10^{-4})$ & -2~(7) \\
 $b\to \rho^-u$ & $4.2\times 10^{-4}~(3.5\times 10^{-4})$ & -2~(7)  \\
 $b\to \pi^0d$ & $5.3\times 10^{-7}~(2.4\times 10^{-6})$ & 93~(31)  \\
 $b\to \rho^0d$ &  $1.4\times 10^{-6}~(5.9\times 10^{-6})$ & 91~(33)  \\
 $b\to \omega\,d$ & $2.5\times 10^{-6}~(5.8\times 10^{-6})$ & -97~(34)  \\
 $b\to\phi\,d$ & $6.9\times 10^{-8}~(2.3\times 10^{-7})$ & -2~(0)  \\
 $b\to\pi^- c$ & $2.2\times 10^{-2}$ & 0 \\
 $b\to\rho^- c$ & $5.1\times 10^{-2}$ & 0 \\
 $b\to\eta\,d$ & $1.5\times 10^{-6}$ & -59 \\
 $b\to\eta' d$ & $1.0\times 10^{-6}$ & 38 \\
 $b\to K^0s$ & $4.0\times 10^{-6}~(2.5\times 10^{-6})$ & -20~(4)   \\
 $b\to K^{*0}s$ & $2.6\times 10^{-6}~(2.9\times 10^{-6})$ & -24~(14)   \\
 $b\to K^-u$ & $9.2\times 10^{-5}~(2.9\times 10^{-5})$ & 5~(28)  \\
 $b\to K^{*-}u$ & $4.8\times 10^{-5}~(5.1\times 10^{-5})$ & 17~(44)  \\
 $b\to \bar K^0d$ & $1.0\times 10^{-4}~(2.0\times 10^{-5})$ & 0.8~(1)  \\
 $b\to \bar K^{*0}d$ & $6.6\times 10^{-5}~(2.6\times 10^{-5})$ & 0.9~(3)   \\
 $b\to K^- c$ & $1.7\times 10^{-3}$ & 0 \\
 $b\to K^{*-}c$ & $2.7\times 10^{-3}$ & 0 \\
 $b\to\eta\,s$ & $1.9\times 10^{-5}$ & -4 \\
 $b\to\eta' s$ & $5.4\times 10^{-5}$ & 1 \\
 $b\to\pi^0s$ & $1.8\times 10^{-6}~(1.6\times 10^{-6})$ & 19~(0)  \\
 $b\to\rho^0s$ & $5.1\times 10^{-6}~(4.3\times 10^{-6})$ & 19~(0)  \\
 $b\to\omega\,s$ & $3.3\times 10^{-7}~(1.3\times 10^{-6})$ & 61~(0)  \\
 $b\to\phi\,s$ & $5.5\times 10^{-5}~(6.3\times 10^{-5})$ & 1~(0)  \\
 $b\to\J\, s$ & $5.4\times 10^{-4}$ & -0.5  \\
 $b\to\J\, d$ & $2.8\times 10^{-5}$ & 10  \\
 \hline
\end{tabular}
\end{center}
\end{table}

For the prompt $\eta'$ production in semi-inclusive decays, we
find the four-quark operator contributions to $b\to\eta's$ can
only account for about 10\% of the measured result \cite{etapK}:
${\cal B}(B\to\eta' X_s)=(6.2\pm 1.6\pm 1.3^{+0.0}_{-1.5}({\rm
bkg}))\times 10^{-4}$ ${\rm for}~2.0<p_{\eta'}<2.7~{\rm GeV/c}$,
where $X_s$ is the final state containing a strange quark. One
important reason is that there is an anomaly effect in the matrix
element $\la\eta'|\bar s\gamma_5s|0\ra$ manifested by the decay
constant $f_{\eta'} ^u$.  As a result, the decay rate of
$b\to\eta's$ induced by the $(S-P)(S+P)$ penguin interaction is
suppressed by the QCD anomaly effect. If there were no QCD
anomaly, one would have ${\cal B}(b\to\eta's)=2.2\times 10^{-4}$
from four-quark operator contributions which are about one third
of the experimental value.

It is known that it proves to be useful to explicitly take into
account the constraints from the CPT theorem when computing PRA's
for inclusive decays at the quark level \cite{Hou} (for a review,
see \cite{Atwood}). The implication of the CPT theorem for partial
rate asymmetries (PRA's) at the hadron level in exclusive or
semi-inclusive reactions is however more complicated
\cite{AtwoodCPT}. Consider the example $b\to du\bar u$. The
corresponding semi-inclusive decays of the $b$ quark can be
manifested as $b\to (\pi^-,\rho^-)u$ and $b\to
(\pi^0,\rho^0,\omega)d$ at the two-body level and $(\pi^-\pi^0,K^0
K^-)u$, $(\pi^+\pi^-,\pi^0\pi^0,K^+ K^-)d$ at the three-body level
and etc. The CPT theorem no longer constrains the absorptive cut
from the $u$-loop penguin diagram not to contribute separately to
each aforementioned semi-inclusive $b$ decay, though the
cancellation between $u\bar u$ and $c\bar c$ quarks will occur
when all semi-inclusive modes are summed over. In view of this
observation, we shall keep all the strong phases in the
calculation of direct $CP$ violation in the individual
semi-inclusive decay.

\section{Semi-inclusive $B$ Decays}
Comparing to two-body decays of the $b$ quark, there exist two
more complications for semi-inclusive $B$ decays. First,  $B\to
MX$ can be viewed as the two-body decay $b\to Mq$ in the heavy
quark limit. For the finite $b$ quark mass, it becomes necessary
to consider the initial $b$ quark bound state effect. Second,
consider the 3-body decay $\ov B\to Mq_1\bar q_2$ with the quark
content $(b\bar q_2)$ for the $\bar B$ meson. One needs a hard
gluon exchange between the spectator quark $\bar q_2$ and the
meson $M$ in order to ensure that the outgoing $\bar q_2$ is hard.
For the semi-inclusive case at hand, it has been argued that the
hard spectator interaction is subject to a phase-space suppression
since it involves three particles in the final states rather than
the two-body one \cite{Hephi}. However, we shall see below that it
is not the case for color-suppressed decay modes, though hard
spectator interactions are formally power suppressed in the heavy
quark limit.

\subsection{Initial bound state effect}
The initial bound state effects on branching ratios and $CP$
asymmetries have been studied recently in \cite{He} using two
different approaches: the light-cone expansion approach and the
heavy quark effective theory approach. We will follow \cite{He} to
employ the second approach.  The nonperturbative HQET parameter
$\mu_G^2$ is fixed from the $B^*-B$ mass splitting to be
$0.36\,{\rm GeV}^2$. Following \cite{He} we use
$\mu_\pi^2=0.5\,{\rm GeV}^2$, which is consistent with QCD sum
rule and lattice QCD calculations \cite{lambda2}. Compared to the
two-body decays $b\to Mq$ shown in Table I, we see that the
branching ratio of $B\to PX$ and $B\to VX$ owing to bound state
effects is reduced by a factor of $(5\sim 10)\%$ and $17\%$,
respectively, while the CP asymmetry remains intact for $VX$
decays and for most of $PX$ modes.

\subsection{Nonfactorizable hard spectator interactions}
We now turn to the hard spectator interactions in the 3-body decay
$B(p_B)\to M(p_M)+q_1(p_1)+\bar q_2(p_2)$ with a hard gluon
exchange between the spectator quark $\bar q_2$ and the meson $M$.
As stressed in Sec.II, the validity of the free $b$ quark
approximation as implied by the parton model indicates that the
hard spectator interaction in semi-inclusive decays is power
suppressed in the heavy quark limit. Using the power counting
$f_B\sim (\Lqcd)^{3/2}/m_b^{1/2}$, $f_M\sim \Lqcd$ \cite{BBNS1},
$\bar\rho\sim \Lqcd/m_b$ and taking into account the phase-space
correction, it is easily seen that the hard spectator interaction
is of order $\Lqcd/m_b$ in the heavy quark limit. However for the
color-suppressed modes such as $\ov B_s\to\pi^0 X_{\bar s}$ in
which the factorizable contribution is color suppressed, the hard
spectator interaction will become extremely important as it is
color allowed.

It is known that the spectator interaction in $B\to K\pi$ decay,
for example, is dominated by soft gluon exchange between the
spectator quark and quarks that form the emitted kaon
\cite{BBNS1}, indicating that QCD factorization breaks down at
twist-3 order. This infrared divergent problem does not occur in
the semi-inclusive decay, however \cite{CS}.

\subsection{Results and discussions}
The results of calculations are shown in Table I. We see that the
tree-dominated color-suppressed modes
$(\pi^0,\rho^0,\omega)X_{\bar s},~\phi X$, $\J X_s,\J X$ and the
penguin-dominated mode $\omega X_{s\bar s}$ are dominated by the
hard spectator corrections. In particular, the prediction ${\cal
B}(B\to \J X_s)=9.6\times 10^{-3}$ is in agreement with the
measurement of a direct inclusive $\J$ production: $(8.0\pm
0.8)\times 10^{-3}$ by CLEO \cite{JX} and $(7.89\pm 0.10\pm
0.34)\times 10^{-3}$ by BaBar \cite{Babar}. This is because the
relevant spectator interaction is color allowed, whereas the
two-body semi-inclusive decays for these modes are
color-suppressed. As a consequence, nonfactorizable hard spectator
interactions amount to giving $a_2$ a large enhancement. In this
work we found that it is the same spectator mechanism responsible
for the enhancement observed in semi-inclusive decay $B\to \J
X_s$, and yet we do not encounter the same infrared problem as
occurred in the exclusive case, and terms proportional to $m_B$
are not power suppressed, rendering the present prediction more
reliable and trustworthy. It is conceivable that infrared
divergences residing in exclusive decays will be washed out when
all possible exclusive modes are summed over.

\begin{table}[h]
\caption{$CP$-averaged branching ratios and partial-rate
asymmetries for some semi-inclusive hadronic $B$ decays with
$E_M>2.1$ GeV for light mesons and $E_\J>3.3$ GeV for the $\J$.
Branching ratios due to hard spectator interactions in the 3-body
decay $B\to Mq_1\bar q_2$ are shown in parentheses. Here $X$
denotes a final state containing no (net) strange or charm
particle, and $X_q$ the state containing the quark flavor $q$.}
\begin{center}
\begin{tabular}{ l l c  }
 Mode & BR & PRA(\%)  \\ \hline
 $\ov B^0\to \pi^-X~(\ov B^0_s\to \pi^-X_{\bar s})$ & $1.3\times 10^{-4}~(5.1\times 10^{-8})$ & -2  \\
 $\ov B^0\to \rho^-X~(\ov B^0_s\to \rho^-X_{\bar s})$ & $3.4\times 10^{-4}~(2.2\times 10^{-7})$ & -2  \\
 $\ov B^0_s\to \pi^0X_{\bar s}$ & $1.3\times 10^{-6}~(8.7\times 10^{-7})$ & 31  \\
 $\ov B^0_s\to \rho^0X_{\bar s}$ &  $4.8\times 10^{-6}~(3.7\times 10^{-6})$ & 22  \\
 $\ov B^0_s\to \omega\,X_{\bar s}$ & $5.5\times 10^{-6}~(3.4\times 10^{-6})$ & -37  \\
 $B^-\to\phi\,X$ & $2.5\times 10^{-7}~(1.9\times 10^{-7})$ & -0.5  \\
 $\ov B^0\to\pi^-X_c~(\ov B_s\to \pi^-X_{c\bar s})$ & $1.8\times 10^{-2}~(8.4\times 10^{-6})$ & 0 \\
 $\ov B^0\to\rho^- X_c~(\ov B_s\to \rho^-X_{c\bar s})$ & $4.2\times 10^{-2}~(1.1\times 10^{-4})$ & 0 \\
 $B^-\to K^0X_s$ & $3.8\times 10^{-6}~(2.9\times 10^{-9})$ & -20   \\
 $B^-\to K^{*0}X_s$ & $2.2\times 10^{-6}~(1.1\times 10^{-8})$ & -24   \\
 $\ov B^0\to K^-X~(\ov B_s\to K^- X_{\bar s})$ & $8.7\times 10^{-5}~(3.6\times 10^{-9})$ & 5   \\
 $\ov B^0\to K^{*-}X~(\ov B_s\to K^{*-} X_{\bar s})$ & $3.9\times 10^{-5}~(1.4\times 10^{-8})$ & 16  \\
 $B^-\to \ov K^0X$ & $9.7\times 10^{-5}~(7.5\times 10^{-8})$ & 0.8  \\
 $B^-\to \ov K^{*0}X$ & $5.4\times 10^{-5}~(2.9\times 10^{-7})$ & 0.9   \\
 $\ov B^0\to K^- X_c~(\ov B^0_s\to K^- X_{c\bar s})$ & $1.4\times 10^{-3}~(4.3\times 10^{-7})$ & 0 \\
 $\ov B^0\to K^{*-}X_c~(\ov B^0_s\to K^{*-} X_{c\bar s})$ & $2.3\times 10^{-3}~(4.8\times 10^{-6})$ & 0 \\
 $\ov B^0_s\to\pi^0X_{s\bar s}$ & $1.5\times 10^{-6}~(5.0\times 10^{-8})$ & 19  \\
 $\ov B^0_s\to\rho^0X_{s\bar s}$ & $4.4\times 10^{-6}~(2.2\times 10^{-7})$ & 18  \\
 $\ov B^0_s\to\omega\,X_{s\bar s}$ & $7.4\times 10^{-6}~(7.1\times 10^{-6})$ & 2  \\
 $B^-\to\phi\,X_s$ & $5.8\times 10^{-5}~(2.8\times 10^{-6})$ & 1  \\
 $B\to\J\, X_s$ & $9.6\times 10^{-3}~(9.2\times 10^{-3})$ & 0 \\
 $B\to\J\, X$ & $5.1\times 10^{-4}~(4.9\times 10^{-4})$ & 0.5  \\
 \hline
\end{tabular}
\end{center}
\end{table}

It is also interesting to notice that after including the
spectator corrections, the branching ratios and PRA's for the
color-suppressed modes $\ov B^0_s\to(\pi^0,\rho^0,\omega)X_{\bar
s}$, $B^-\to\phi X$ are numerically close to that predicted in
\cite{Soni} based on naive factorization (see Table I). Note that
the large CP asymmetries in $b\to (\pi^0,\rho^0,\omega)d$ decays
(see Table I) are washed out to a large extent at hadron level by
spectator interactions. By contrast, the nonfactorizable spectator
interaction is in general negligible for penguin dominated (except
for $\omega X_{s\bar s}$) or color-allowed tree dominated decay
modes. The channels $(B^-,\ov B^0)\to(\pi^0,\rho^0,\omega,\phi)X$
are not listed in Table II as they involve the unwanted form
factors. For example, $B^-\to\pi^0X$ contains a term $a_2F^{B\pi}$
and $\ov B^0\to\pi^0X$ has a contribution like $a_4F^{B\pi}$.
Hence, the prediction of $(B^-,\ov B^0)\to\pi^0X$ is not as clean
as $\ov B_s^0\to \pi^0X_{\bar s}$. Nevertheless, the former is
also dominated by spectator interactions and is expected to have
the same order of magnitude for branching ratios as the latter.

Owing to the presence of $B-\eta(\eta')$ form factors, the decays
$B\to (\eta,\eta')X$ are also not listed in Table II.  However, we
find that the hard spectator corrections to the prompt $\eta'$
production in semi-inclusive decays are very small and hence the
four-quark operator contributions to $b\to\eta' s$ can only
account for about 10\% of the measured result. Evidently this
implies that one needs a new mechanism (but not necessarily new
physics) specific to the $\eta'$. It has been advocated that the
anomalous coupling of two gluons and $\eta'$ in the transitions
$b\to sg^*$ followed by $g^*\to \eta'g$ and $b\to sg^*g^*$
followed by $g^*g^*\to\eta'$ may explain the excess of the $\eta'$
production \cite{AS,Kagan}. An issue in this study is about the
form-factor suppression in the $\eta'-g^*-g^*$ vertex and this has
been studied recently in the perturbative QCD hard scattering
approach \cite{Ali00}. At the exclusive level, it is well known
that the decays $B^\pm\to\eta'K^\pm$ and $\ov B^0\to\eta' \ov K^0$
have abnormally large branching ratios \cite{PDG}. In spite of
many theoretical uncertainties, it is safe to say that the
four-quark operator contribution accounts for at most half of the
experimental value and the new mechanism responsible for the
prolific $\eta'$ production in semi-inclusive decay could also
play an essential role in $B\to\eta'K$ decay.

From Table II it is clear that the semi-inclusive decay modes:
$\ov B^0_s\to(\pi^0,\rho^0,\omega)X_{\bar s}$, $\rho^0X_{s\bar
s}$, $\ov B^0\to (K^-X,K^{*-}X)$ and $B^-\to(K^0X_s,K^{*0}X_s)$
are the most promising ones in searching for direct CP violation;
they have branching ratios of order $10^{-6}-10^{-4}$ and CP rate
asymmetries of order $(10-40)\%$. Note that  a measurement of
partial rate difference of $\ov
B^0_s\to(\pi^0,\rho^0,\omega)X_{\bar s}$ and
$B^-\to(K^0X_s,K^{*0}X_s)$ will provide useful information on the
unitarity angle $\alpha$, while $\ov B^0_s\to\rho^0X_{s\bar s}$
and $\ov B^0\to (K^-X,K^{*-}X)$ on the angle $\gamma$. To have a
rough estimate of the detectability of CP asymmetry, it is useful
to calculate the number of $B-\ov B$ pairs needed to establish a
signal for PRA to the level of three statistical standard
deviations given by
 \be
N_B^{3\sigma}={9\over \Delta^2 Br\, \epsilon_{\rm eff}},
\label{NB}
 \en
where $\Delta$ is the PRA, $Br$ is the branching ratio and
$\epsilon_{\rm eff}$ is the product of all of the efficiencies
responsible for this signal. With about $1\times 10^7$ $B\ov B$
pairs, the asymmetry in $K^{*-}$ channel starts to become
accessible; and with about $7\times 10^7$ $B\ov B$ events, the
PRA's in the other modes mentioned above will become feasible.
Here we assumed, for definiteness, $\epsilon_{\rm eff}=1$ and a
statistical significance of 3$\sigma$ as in Eq. (\ref{NB}).
Currently BaBar has collected 23 million $B\ov B$ events, BELLE 11
million pairs and CLEO 9.6 million pairs. It is conceivable that
CP asymmetries in semi-inclusive $B$ decays will begin to be
accessible at these facilities. Likewise, PRA's in semi-inclusive
$B_s$ decays may be measurable in the near future at the
Fermilab's Tevatron.

It is interesting to note that the decays $\ov B^0_s\to
(\pi^0,\rho^0,\omega)X_{s\bar s}$ and $B^-\to\phi X$ are
electroweak-penguin dominated. Except for the last channel, they
have sizable branching ratios and two of them have observable CP
asymmetries. A measurement of these reactions will provide a good
probe of electroweak penguins.

\section{Conclusions}
We have systematically investigated semi-inclusive $B$ decays
$B\to MX$ within a framework inspired by QCD-improved
factorization. The nonfactorizable effects, such as vertex-type
and penguin-type corrections to the two-body $b$ decay, $b\to Mq$,
and hard spectator corrections to the 3-body decay $B\to Mq_1\bar
q_2$ are calculable in the heavy quark limit. QCD factorization
seems applicable when the emitted meson is a light meson or a
charmonium.

There are two strong phases in the QCD factorization approach: one
form final-state rescattering due to hard gluon exchange between
$M$ and $X$, and the other from the penguin diagrams. The strong
phase coming from final-state rescattering due to hard gluon
exchange between the final states $M$ and $X$ can induce large
rate asymmetries for tree-dominated color-suppressed modes
$(\pi^0,\rho^0,\omega)X_{\bar s}$. The predicted coefficient $a_2$
in QCD factorization is very small compared to naive
factorization. Consequently, the color-suppressed modes
$(\pi^0,\rho^0,\omega)X_{\bar s},~\phi X$ and $\J X_s,\J X$ are
very suppressed. Fortunately, the nonfactorizable hard spectator
interactions in $B\to Mq_1\bar q_2$, though phase-space
suppressed, are extremely important for the aforementioned modes.
Our prediction ${\cal B}(B\to \J X_s)=9.6\times 10^{-3}$ is in
agreement with experiment. Contrary to the exclusive hadronic
decay, the spectator quark corrections here are not subject to the
infrared divergent problem, rendering the present prediction more
clean and reliable.

$\ov B^0_s\to(\pi^0,\rho^0,\omega)X_{\bar s}$, $\rho^0X_{s\bar
s}$, $\ov B^0\to (K^-X,K^{*-}X)$ and $B^-\to(K^0X_s,K^{*0}X_s)$
are the most promising semi-inclusive decay modes in searching for
direct CP violation; they have branching ratios of order
$10^{-6}-10^{-4}$ and CP rate asymmetries of order $(10-40)\%$.
With about $7\times 10^7$ $B\ov B$ pairs, CP asymmetries in these
modes may be measurable in the near future at the BaBar, BELLE,
CLEO and Tevatron experiments. The decays $\ov B_s^0\to
(\pi^0,\rho^0,\omega)X_s$ and $B^-\to\phi X$ are
electroweak-penguin dominated. Except for the last mode, they in
general have sizable branching ratios and two of them have
observable CP asymmetries. The above-mentioned reactions will
provide good testing ground for the standard model and a good
probe for electroweak penguins.

\section*{Acknowledgments} I wish to thank Yue-Laing Wu for organizing this wonderful
conference and Amarjit Soni for collaboration on this interesting
topic. I would also like to thank Physics Department, Brookhaven
National Laboratory for its hospitality.



\section*{References}
\begin{thebibliography}{99}
\newcommand{\bi}{\bibitem}

\bi{Soni} D. Atwood and A. Soni, {\PRL} {\bf 81}, 3324 (1998).

\bi{Browder} T.E. Browder, A. Datta, X.G. He, and S. Pakvasa,
{\PRD} {\bf D57}, 6829 (1998).

\bi{BBNS1} M. Beneke, G. Buchalla, M. Neubert, and C.T. Sachrajda,
{\PRL} {\bf 83}, 1914 (1999); {\NPB} {\bf 591}, 313 (2000);
hep-ph/0104110.

\bi{He} X.G. He, C. Jin, and J.P. Ma, {\PRD} {\bf 64}, 014020
(2001).

\bi{Hephi} X.G. He, J.P. Ma, and C.Y. Wu, \PRD {\bf 63}, 094004
(2001).

\bi{CS} H.Y. Cheng and A. Soni, hep-ph/0105246, to appear in Phys.
Rev. D.

\bi{etapK} CLEO Collaboration, T.E. Browder {\it et al.,} {\PRL}
{\bf 81}, 1786 (1998).

\bi{Hou} J.M. G\'erard and W.S. Hou, {\PRD} {\bf 43}, 2909 (1991).

\bi{Atwood} D. Atwood, S. Bar-Shalom, G. Eilam, and A. Soni, {\em
Phys. Rep.} {\bf 347}, 1 (2001).


\bi{AtwoodCPT} D. Atwood and A. Soni, {\PRD} {\bf 58}, 036005
(1998).

\bi{lambda2} P. Ball and V.M. Braun, {\PRD} {\bf 49}, 2472 (1994);
A.S. Kronfeld and J.N. Simone {\PLB} {\bf 490}, 228 (2000).

\bi{JX} CLEO Collaboration, R. Balest {\it et al.,} {\PRD} {\bf
52}, 2661 (1995).

\bi{Babar} BaBar Collaboration, V. Brigljevic, invited talk
presented at XXXVI Recontres de Moriond on QCD, March 17-24, 2001.


\bi{AS} D. Atwood and A. Soni, {\PLB} {\bf 405}, 150 (1997);
{\PRL} {\bf 79}, 5206 (1997); W.S. Hou and B. Tseng, {\PRL} {\bf
80}, 434 (1998); X.G. He and G.L. Lin, {\PLB} {\bf 454}, 123
(1999); M.R. Ahmady, E. Kou, and A. Sugamoto, {\PRD} {\bf 58},
014015 (1998); D. Du, C.S. Kim, and Y. Yang, {\PLB} {\bf 426}, 133
(1998).

\bi{Kagan} A.L. Kagan and A.A. Petrov, UCHEP-27 [hep-ph/9707354].

\bi{Ali00} A. Ali and A.Ya. Parkhomenko, hep-ph/0012212.

\bi{PDG} Particle Data Group, D.E. Groom {\it et al.,} {\sl Eur.
Phys. J}, {\bf C15}, 1 (2000).




\end{thebibliography}
\end{document}